\documentclass[preprint,aps,pra]{revtex4}
\usepackage{epsf}

\newcommand{\Btwop}{$\beta_2^{(+)}$}
\newcommand{\Btwom}{$\beta_2^{(-)}$}
\newcommand{\Bfourp}{$\beta_4^{(+)}$}
\newcommand{\Bfourm}{$\beta_4^{(-)}$}

\begin{document}
\title{Anisotropy parameters in two-colour two-photon above threshold ionization}

\author{Steven Hutchinson}
\email{steven.hutchinson@ucl.ac.uk}
\affiliation{Department of Physics and Astronomy, University College London, Gower Street, London, WC1E 6BT, United Kingdom}
\author{Michael A. Lysaght}
\affiliation{Irish Centre for High End Computing,
Tower Building, Trinity Technology and Enterprise Campus,
Grand Canal Quay,
Dublin 2,
Ireland} 
\author{Hugo W. van der Hart}
\affiliation{Centre for Theoretical Atomic, Molecular and Optical Physics, School of
Mathematics and Physics, Queen's University
Belfast, Belfast BT7 1NN, United Kingdom}

\date{\today}

\begin{abstract}
We employ the time-dependent R-Matrix (TDRM) method to calculate anisotropy parameters
for positive and negative sidebands of selected harmonics generated by two-color two-photon
above threshold ionization of argon. We consider odd harmonics of an 800 nm field ranging
from the 13$^\textrm{th}$ to 19$^\textrm{th}$ harmonic, overlapped by a fundamental
800 nm IR field. The anisotropy parameters obtained using the TDRM method are compared
with those obtained using a second order perturbation theory with a model potential (MP)
approach and a soft photon approximation (SPA) approach. Where available, a comparison
is also made to published experimental results. All three theoretical approaches provide
similar values for anisotropy parameters. The TDRM approach obtains values that are
closest to published experimental values. At high photon energies, the differences
between each of the theoretical methods become less significant.
\end{abstract}

\maketitle

\section{Introduction}

In recent years, one of the most prominent areas of atomic physics has been the study of atomic
processes on an ultra-fast timescale \cite{Krausz09}. This work has been driven by the development
of ultra-fast light sources capable of producing light pules with a duration in the attosecond
region \cite{Kienberger04}, which has enabled real-time experimental observation of ultra-fast
atomic behavior such as light-field-induced electron tunneling \cite{Uiberacker07}, the decay
of an inner-shell vacancy \cite{Drescher02} and the motion of a valence electron \cite{Goulielmakis10}
while attosecond electron wavepacket interferometry has revealed information about the ultrafast
dynamics of electron wavepackets \cite{Remetter06}. Key to sustaining these developments in pulse
generation and application has been the characterization of light pulses of such short duration.

The characterization of attosecond pulses has proven to be highly challenging, as light pulses on
the attosecond timescale have relatively low intensity. This renders most standard short-pulse
characterization techniques impossible, since these tend to be based on intensity autocorrelation
in non-linear processes \cite{Walmsley09}. For example, techniques using nonlinear crystals such
as FROG \cite{Trebino97} cannot be used (directly) since these crystals are highly absorbent in
the XUV range. This has led to the development of new metrology methods which use the non-linearity
of processes such as high harmonic generation and multi-color two-photon ionization for attosecond
pulse characterization. One of the most successful of these has been the reconstruction of attosecond
beating by interference of two-photon transition (RABBITT) \cite{Paul01}. This method uses three-color
two-photon ionization to generate two consecutive high harmonic signals and three associated sidebands.
By observing the modulation of the central common sideband relative to the time at which an IR field
is applied, it is possible to determine the relative phase of each of the original harmonic pulses.
Once the relative phase of all components has been obtained, the original pulse can be reconstructed
\cite{Aseyev03}.

Due to the importance of ATI sidebands for pulse characterization, there has been recent experimental
interest in photoelectron angular distributions of individual ATI sidebands in two-color two-photon
ionization processes. In particular, experimental measurements have recently been made to determine
the photoelectron angular distributions of positive and negative sidebands of the 13$^{\textrm{th}}$
and 15$^{\textrm{th}}$ harmonics of Argon in two-color two-photon above threshold ionization \cite{Haber09}.
In this study, experimental measurements were used to determine the anisotropy parameters of each sideband
and the ratio between cross sections for positive and negative sidebands for each harmonic. A comparison
was then made between experimental values and the values obtained from theoretical approaches using
second order perturbation theory with a model potential (MP) \cite{Toma02} and the soft photon
approximation (SPA) \cite{Maquet07}. Although there was reasonable agreement with experimental
values, both of the theoretical models produced anisotropy parameters outside of the range of
experimental error.

The discrepancy between experimental and theoretical results displayed in \cite{Haber09} creates an
interest to apply other theoretical methods which account for more of the atomic structure to this
type of problem. Both the model potential and the soft photon approximation methods are based on
the single active electron (SAE) approximation. This approximation significantly reduces the
computational complexity of modeling harmonic generation but cannot describe the full physics
involved which may play a role on ultra-short timescales. Central to the MP approach \cite{Toma02}
is the use of a model potential in order attempt to compensate for electron exchange and correlation
effects which may provide a significant limitation on the accuracy of this approach. It is therefore
of interest to investigate how sensitive these asymmetry parameters are with respect to the potential,
by comparing with results from, for example, an ab-initio approach.

Theoretical approaches that numerically solve the full-dimensional time-dependent Schr\"{o}dinger equation
for multielectron systems in a laser field with exact potentials acting on each of the electrons are
available \cite{Smyth19981-HELIUM}, however the complexity of such a system means that the problem is
intractable for targets with more than two active electrons. The recently developed time dependent
R-matrix (TDRM) theory \cite{Lysaght09}, however, provides a technique capable of describing the
time-dependent response of a general multielectron system interacting with a laser field while
employing R-matrix methods such as space partition to reduce the computational complexity. At
the moment this method is limited to the emission of a single electron. Unlike the MP and SPA
methods, the TDRM method uses the real potential acting on each electron as multielectron effects
such as electron correlation and exchange are properly accounted for.

The TDRM method has already proven highly successful in applications to the study of ultra-fast
electron dynamics \cite{Lysaght09.C+ion,Lysaght08,Lysaght09.C+momentum}. As sidebands in two-color
two-photon above threshold ionization are of significant experimental \cite{Haber09} and theoretical
\cite{Veniard96} interest, with previous comparable work having already been performed \cite{Haber09},
it is of interest to apply the TDRM method to two-color two-photon above threshold ionization of Argon.
We use the same laser frequencies from this experiment to calculate anisotropy parameters for positive
and negative sidebands, and the ratio between the cross sections of these sidebands, of the
13$^{\textrm{th}}$ and 15$^{\textrm{th}}$ harmonics using the TDRM method to enable a comparison
with experimentally measured values. We also compare the previously applied MP \cite{Toma02} and
SPA \cite{Maquet07} theoretical methods by verifying anisotropy parameters and cross section ratios
for relevant sidebands using these methods. Finally, we extend our results beyond those that have
been experimentally measured to include the 17$^{\textrm{th}}$ and 19$^{\textrm{th}}$ harmonics
using the TDRM method, and, where available, the SPA and MP methods.

\section{Time Dependent R-Matrix Theory}

The time dependent R-matrix theory used throughout this study is an extension of standard
R-matrix techniques for scattering processes to time-dependent processes. A thorough overview
of this theory has been published previously \cite{Lysaght09}, thus only a brief description
is given here.

The TDRM method solves the time-dependent Schr\"{o}dinger equation for a general $(N+1)$ electron
atom or ion interacting with a laser pulse by employing the unitary form of the time evolution
operator to rewrite the TDSE in the form of a Crank-Nicolson scheme as follows:
\begin{equation}
\label{eq:AR CN1}
[H(t_{q+1/2})-E]\Psi(\mathbf{X}_{N+1},t_{q+1})=\Theta(\mathbf{X}_{N+1},t_{q}),
\end{equation}
where
\begin{equation}
\label{eq:AR CN2}
\Theta(\mathbf{X}_{N+1},t_{q})=-[H(t_{q+1/2})+E]\Psi(\mathbf{X}_{N+1},t_{q}).
\end{equation}
In equations (\ref{eq:AR CN1}) and (\ref{eq:AR CN2}),
$\mathbf{X}_{N+1}=\mathbf{x}_1,\mathbf{x}_2,\ldots,\mathbf{x}_{N+1}$ where
$\mathbf{x}_i\equiv\mathbf{r}_i\sigma_i$ are the space and spin coordinates
of the $i$th electron. To implement the Crank-Nicolson scheme, we have
introduced a discrete mesh in time with a discrete time step $\Delta t= t_{q+1}-t_q$.
The imaginary energy $E$ is then defined by this time step according to
$E \equiv 2i(\Delta t)^{-1}$. $H(t_{q+1/2})$ represents the time-dependent Hamiltonian
at the midpoint of times $t_q$ and $t_{q+1}$. We assume that the light field is
spatially homogeneous and linearly polarized throughout. Following the analysis
presented in earlier work \cite{Hutchinson10}, which demonstrated that the optimum
choice of gauge for this type of problem was the length gauge, the length gauge is
used to describe the laser interaction throughout.

To solve equation (\ref{eq:AR CN1}), we employ standard R-matrix techniques by partitioning the
configuration space into two distinct regions: an internal region and an external region. The
internal region is defined as a small region with radius $r=a_{\textrm{in}}$ chosen to
enclose the core of the target, with all $(N+1)$ electrons contained within this region.
In the internal region exchange and correlation effects are considered between all of the
$(N+1)$ electrons. The external region is defined as a large spatial region
$a_{\textrm{in}} \leq r \leq a_{\textrm{out}}$ where only the ejected $(N+1)^\textrm{th}$
electron is present. The residual $N$ electrons are still considered, however they are
confined to the internal region spatially with correlation effects accounted for by long-range
potential matrices. Exchange effects between the ejected electron and the residual $N$
electrons are considered negligible and thus not included. The external region is chosen
with $a_{\textrm{out}}$ large enough that the ejected electron wavefunction does not
reach this boundary within the finite time considered. For computational reasons the
external region is further subdivided into subregions of identical length.

In the internal region we expand the wavefunction $\Psi(\mathbf{X}_{N+1},t_{q+1})$ in
an antisymmetric R-matrix basis, $\psi_k$. To ensure Hermicity at the boundary
$r=a_{\textrm{in}}$, and to account for the component of the wavefunction that
leaves the internal region box, we introduce a Bloch operator $\mathcal{L}$ which
allows us to rewrite equation (\ref{eq:AR CN1}) in the form:
\begin{equation}
\label{eq:Ar main ham}
(H+\mathcal{L}-E)\Psi_{q+1}=\mathcal{L}\Psi_{q+1}+\Theta_q,
\end{equation}
which has the formal solution
\begin{equation}
\label{eq:Ar formal sol}
\Psi=(H+\mathcal{L}-E)^{-1}\mathcal{L}\Psi+(H+\mathcal{L}-E)^{-1}\Theta.
\end{equation}
Solutions of this equation are found by expressing the wavefunction $\Psi$ in the inner
region in terms of inner region eigenfunctions $\psi_k$ of the operator $(H+\mathcal{L})$:
\begin{equation}
\label{eq:Ar Bk exp}
\Psi(\mathbf{X}_{N+1},t_{q+1})=\sum_k \psi_k(\mathbf{X}_{N+1})B_k(E,t_{q+1}),
\end{equation}
where $B_k$ are time-dependent expansion coefficients.
To connect the internal region and the external region, we first consider the behavior
of the internal region wavefunction at the boundary between regions. To this end, we
project equation (\ref{eq:Ar formal sol}) onto the $n$ time-independent channel
functions $\bar{\Phi}_p^\gamma$, which are formed by coupling the residual ion
state $\Phi$ with the angular and spin functions of the continuum electron. By evaluating
the resulting expression on the boundary $r=a_{\textrm{in}}$ we obtain the following expression:
\begin{equation}
\label{eq:Ar RTF}
\mathbf{F}(a_{\textrm{in}})=\mathbf{R}a_{\textrm{in}}\mathbf{\bar{F}}(a_{\textrm{in}})
+\mathbf{T}(a_{\textrm{in}}),
\end{equation}
where $\mathbf{F}$ is the reduced radial wavefunction and $\mathbf{\bar{F}}$ its first
derivative. The terms $\mathbf{R}$ and $\mathbf{T}$ represent the R-matrix and T-vector
respectively. Formal definitions for each of these terms are available in
reference \cite{Lysaght09}. The right hand side of equation (\ref{eq:Ar RTF}) consists
of two main components. The T-vector arises from the action of the operator
$(H+\mathcal{L}-E)^{-1}$ on the inhomogeneous term $\Theta$ in equation (\ref{eq:Ar formal sol})
and provides information about the flow of the wavefunction at $t=t_q$ through the boundary.
The $\mathbf{R}a_{\textrm{in}}\mathbf{\bar{F}}(a_{\textrm{in}})$ term is a correction to
account for the components of the wavefunction that leave or enter the internal region,
and thus provides information about the rate of flow of the unknown wavefunction at
$t=t_{q+1}$ through the boundary. This term arises from the action of the Bloch operator
in equation (\ref{eq:Ar formal sol}). By obtaining the vector $\mathbf{F}$ we may
determine the expansion coefficients $B_k$ in equation (\ref{eq:Ar Bk exp}) and
consequently the full wavefunction $\Psi$ in the internal region at $t=t_{q+1}$.
However, we must first determine the modified derivative functions $\mathbf{\bar{F}}$
from analysis of the external region.

In the external region we expand the wavefunction according to
\begin{equation}
\label{eq:Ar ext exp}
\Psi(\mathbf{X}_{N+1},t_{q+1})=\sum_{p=1}^n \bar{\Phi}_p^\gamma
(\mathbf{X}_N;\mathbf{\hat{r}}_{N+1})r_{N+1}^{-1}F_p(r_{N+1}),
\end{equation}
where the reduced radial functions $F_p$ are analytic continuations
of the functions defined on the internal region boundary in equation (\ref{eq:Ar RTF}).
As in the internal region, we introduce a Bloch operator to ensure hermicity on the boundaries
of each of the subregions in the external region. The formal solution to equation (\ref{eq:AR CN1})
in the external region then has the same form as equation (\ref{eq:Ar formal sol}). The
expansion (\ref{eq:Ar ext exp}) is such that, by using similar techniques to those used
in the internal region, we may then demonstrate that equation (\ref{eq:Ar RTF}) is valid
for any boundary between subregions in the external region.

Having validated equation (\ref{eq:Ar RTF}) for any boundary, we may now develop an approach to
determine the wavefunction $\mathbf{F}$ and its derivative $\mathbf{\bar{F}}$ in the
external region. We propagate the R-matrix and T-vector outwards across the boundaries of
each of the external region sectors through the use of Greens functions derived from the
Hamiltonian, using the R-matrix and T-vector calculated on the internal region boundary
$r=a_{\textrm{in}}$ from analysis of the internal region as the initial values. The details
of the propagators used to accomplish this are provided in \cite{Lysaght09}. By choosing the
external region outer boundary large enough to ensure the wavefunction does not reach the
outer limit $r=a_{\textrm{out}}$ in the timeframe considered, we may impose the boundary
condition $\mathbf{F}=0$ at $r=a_{\textrm{out}}$ for every time step. Using the R-matrix
and T-vectors on the boundary of each subregion, we may then propagate this F-vector
inwards to determine its values at every boundary, providing the wavefunction $\Psi$ at
every point of the external region. Finally, we may also determine $\mathbf{\bar{F}}$ on
the internal region boundary, and consequently the wavefunction $\Psi$ at $t=t_{q+1}$ in
the internal region. Having now obtained the wavefunction for the entire configuration
space at $t=t_{q+1}$, we may use this wavefunction as the starting point for the next
iteration of the procedure.

To describe Argon we use the R-Matrix basis developed for single photon ionization of
Ar \cite{Burke75}, which includes the $3s^23p^5$ $^2$P$^o$ and $3s3p^6$ $^2$S$^e$ states
of Ar$^+$ as target states, with all $3s^23p^5\epsilon l$ and $3s3p^6\epsilon l$ channels
with angular momentum up to an including $L_{\textrm{max}}=19$ included in the
description of Argon. The internal region is chosen to extend to a radius of 20 au,
with the set of continuum orbitals containing 70 continuum functions for each available
angular momentum of the continuum electron. The external region is chosen to extend to
a distance of 1826 au and is composed of subsectors of width 2 au which contain 40
B-splines per channel with order $k=11$.

We consider irradiation by an EUV laser pulse overlapped by an IR dressing field. The IR
laser pulse has a wavelength of 800nm with a peak intensity of $5\times10^{10}$ W cm$^{-2}$
and a pulse profile consisting of a 3-cycle sin$^2$ ramp on, followed by 2 cycles at peak
intensity and a 3-cycle sin$^2$ ramp off. The EUV laser pulse corresponds to a selected
harmonic of an 800nm pulse ranging from the 13$^{\textrm{th}}$ to 19$^{\textrm{th}}$ harmonic
and has a peak intensity $5\times 10^{11}$ W cm$^{-2}$. The EUV pulse for the $n^\textrm{th}$
harmonic is described by a $3n$-cycle sin$^2$ ramp on, followed by $2n$ cycles at peak intensity
and a $3n$-cycle sin$^2$ ramp off. The EUV and IR pulses start concurrently and are in phase.

\section{Results}

We apply the TDRM method to calculate the wavefunction of the ejected electron when neutral
Argon is simultaneously irradiated by an EUV pulse corresponding to a selected harmonic of
an 800nm pulse ranging from the 13${^\textrm{th}}$ to the 19${^\textrm{th}}$ harmonic with
intensity $5\times 10^{11}$ W cm$^{-2}$, overlapped by an IR dressing field with wavelength
800nm and intensity $5\times10^{10}$ W cm$^{-2}$. The overlapping laser fields generate
positive and negative sidebands of a central single photon ionization peak. After the
pulses end, the wavefunction is propagated for a further 1.87 fs.

\begin{figure}[htbp]
\centerline{\epsfxsize=6in\epsffile{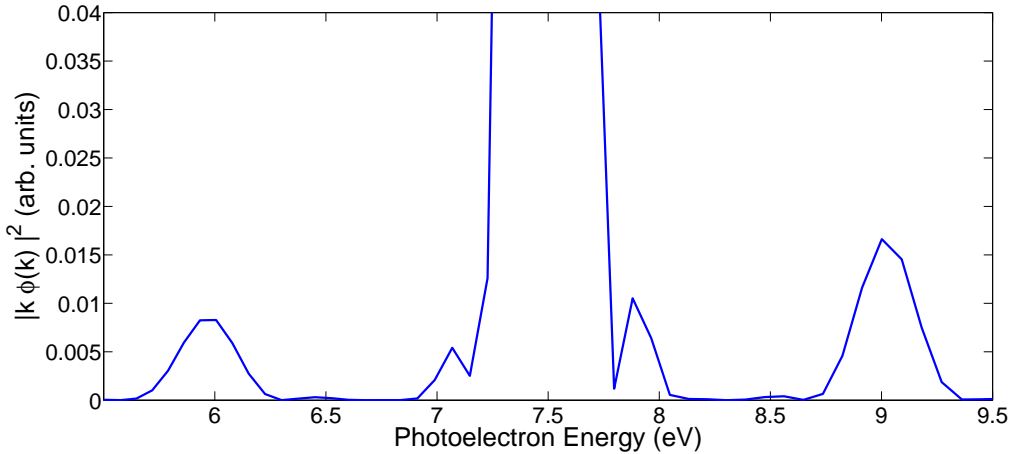}}
\caption{(Color online) Photoelectron energy spectrum along the $y$ axis for the 15$\textrm{th}$ harmonic
and overlapping fundamental of the 800nm pulse for Argon showing the central single photon
ionization peak and associated positive and negative sidebands.}
\label{fig:ArDist}
\end{figure}

In order to obtain the photoelectron angular distribution for the positive and negative sidebands
generated in Argon by a selected harmonic of the 800nm pulse and the overlapping fundamental IR
pulse, we transform the wavefunction of the outer electron 1.87 fs after the end of the laser
pulse to obtain a 2D momentum distribution for the ejected electron as explained previously
\cite{vdHart08}. This transformation assumes that the radial potential has become negligible
in the region of the transformation.
A typical photoelectron energy spectrum, obtained for the 15$^{\textrm{th}}$ harmonic of the
800nm pulse, for the angle $\theta=0$ using the TDRM method is shown in figure \ref{fig:ArDist}.
When compared to the experimentally measured photoelectron energy spectra provided in \cite{Haber09},
we observe that the central harmonic peak is much narrower with the positive and negative sidebands
clearly separated from this main peak. We also note that the intensity of the sidebands is
significantly lower relative to the central harmonic peak than observed in the experimental
results. This may be due to the shorter duration of the current pulses compared to the
experimental pulse, as the central peak corresponds to a single photon process, and as
such has an intensity that scales linearly with pulse length, whereas the sidebands
correspond to a two photon process with an intensity that scales quadratically with
pulse length. This use of a shorter pulse was imposed by computational limits. The
much narrower harmonic peak and distinct sidebands allow us to safely integrate over
the width of each sideband at a given angle to calculate the relevant photoelectron
angular distribution $I(\theta)$ for each sideband.

As the laser fields considered in these calculations are both linearly polarized in the
$z$ direction, and the target is unaligned, the photoelectron angular distributions for
two-photon ionization are known to be of the form \cite{Reid03}:
\begin{equation}
\label{eq:leg_fit}
I(\theta) \propto \frac{\sigma}{4\pi}[1+\beta_2 P_2(\cos \theta) + \beta_4 P_4 (\cos \theta)],
\end{equation}
where $\beta_n$ are the normalized anisotropy parameters.
Having calculated the photoelectron angular distributions $I(\theta)$ for sidebands of high harmonics of
argon using the TDRM method, we fit Legendre polynomials of the form of equation (\ref{eq:leg_fit}) to
the angular distribution to obtain the normalized anisotropy parameters $\beta_n$. The ratio
$\sigma^{(+)}/\sigma^{(-)}$ between the positive and negative sidebands is
obtained by comparing the constant scaling factor when fitting the Legendre polynomials
to each of the sidebands. This ratio should thus be considered a ratio of electron emission yields
rather than a ratio of cross sections. Anisotropy parameters for the SPA and MP methods have been verified
using the same technique, using data generated by the TDRM approach for single-photon ionization
and the tabulated data \cite{Toma02}, respectively.
\begin{center}
\begin{table}
\caption{Anisotropy parameters and cross section ratios for sidebands generated by the 13$^{\textrm{th}}$ to 19$^{\textrm{th}}$ harmonics of the 800nm pulse overlapped by the fundamental IR pulse in Argon}
\label{tab:ArResults}
\begin{tabular}{lccccc}
  \hline
  \hline
   & $\sigma^{(+)}/\sigma^{(-)}$  & $\beta_{2}^{(-)}$ & $\beta_{4}^{(-)}$ & $\beta_{2}^{(+)}$ & $\beta_{4}^{(+)}$ \\
   \hline
  13$^{\textrm{th}}$ HH experiment & 1.39$\pm$0.05 & 1.52$\pm$0.05 & 0.21$\pm$0.05 & 2.37$\pm$0.10 &  0.63$\pm$0.04  \\
  13$^{\textrm{th}}$ HH MP & 1.42 & 2.08 & 0.50 & 2.72 & 1.18 \\
  13$^{\textrm{th}}$ HH SPA (TDRM $\beta$) & 1.75 & 2.32 & 0.43 & 2.52 & 0.70 \\
  13$^{\textrm{th}}$ HH TDRM & 1.53 & 1.87 & 0.28 & 2.32 & 0.67 \\
  \hline
  15$^{\textrm{th}}$ HH experiment & 1.27$\pm$0.05 & 1.63$\pm$0.12 & 0.54$\pm$0.15 & 2.48$\pm$0.06 & 0.69$\pm$0.05 \\
  15$^{\textrm{th}}$ HH MP & 1.28 & 2.42 & 0.72 & 2.87 & 1.27 \\
  15$^{\textrm{th}}$ HH SPA (TDRM $\beta$) & 1.31 & 2.52 & 0.70 & 2.63 & 0.84 \\
  15$^{\textrm{th}}$ HH TDRM & 1.35 & 2.25 & 0.52 & 2.58 & 0.88 \\
  \hline
  17$^{\textrm{th}}$ HH experiment &  &  &  &  & \\
  17$^{\textrm{th}}$ HH MP & 1.22 & 2.57 & 0.84 & 2.92 & 1.28\\
  17$^{\textrm{th}}$ HH SPA (TDRM $\beta$) & 1.10 & 2.63 & 0.84 & 2.69 & 0.92\\
  17$^{\textrm{th}}$ HH TDRM & 1.13 & 2.36 & 0.63 & 2.72 & 1.06 \\
  \hline
  19$^{\textrm{th}}$ HH experiment &  &  &  &  & \\
  19$^{\textrm{th}}$ HH MP &  & 2.64 & 0.91 &  & \\
  19$^{\textrm{th}}$ HH SPA (TDRM $\beta$) &  & 2.69 & 0.92 & & \\
  19$^{\textrm{th}}$ HH TDRM & 1.16 & 2.49 & 0.74 & 2.76 & 1.05 \\
\hline
\hline
\end{tabular}
\end{table}
\end{center}

For odd numbered harmonics ranging from the 13$^{\textrm{th}}$ to 19$^{\textrm{th}}$ harmonic of the
800nm pulse, table \ref{tab:ArResults} presents anisotropy parameters for positive and negative
sidebands and cross section ratios using the MP, SPA and TDRM methods where available. Also provided
for the 13$^{\textrm{th}}$ and 15$^{\textrm{th}}$ harmonics are experimentally measured anisotropy
parameters from \cite{Haber09}. In table \ref{tab:ArResults} anisotropy parameters for negative
sidebands are denoted by superscript $(-)$, and likewise positive sidebands are denoted by
superscript $(+)$. The experimental data in \cite{Haber09} were compared with results from
MP \cite{Toma02} and SPA \cite{Maquet07}. We have therefore included anisotropy parameters
obtained via these approaches as well.

The anisotropy parameters and cross section ratios calculated using the TDRM method show that, for each
of the selected harmonics the negative sideband is smaller in magnitude than the positive sideband.
This trend is observed in all of the results presented in table \ref{tab:ArResults}. We also note,
that as the energy of the harmonic increases the anisotropy parameters increase, but by diminishing
amounts. This is demonstrated particularly clearly for the positive sideband, where values for the
17$^{\textrm{th}}$ and 19$^{\textrm{th}}$ harmonic are highly similar. Significantly, when using
the TDRM method, sidebands with the same energy have differing anisotropy parameters for positive
and negative sidebands, with negative sidebands having lower values than a positive sideband at
the same energy. This demonstrates that there are slight differences in the physics of emission
and absorption processes involving the IR photon.

The values we obtain for the anisotropy parameters using the TDRM method in table
\ref{tab:ArResults} demonstrate varying agreement with experimental values. Qualitatively,
the asymmetry parameters behave in a similar fashion in experiment and theory. The quantitative
agreement between TDRM and experiment is best for positive sidebands with the TDRM method
providing values for the anisotropy parameters of the positive sideband of the 13$^{\textrm{th}}$
high harmonic that lie entirely within the experimental range of values.
There remains a discrepancy however between anisotropy parameters measured experimentally and
those obtained using the TDRM method for the negative sideband. For example the \Btwom parameter
for the 15$^{\textrm{th}}$ harmonic calculated using the TDRM method lies well outside the range
of experimental values. This may be explained in part by differences in the experimental and
theoretical pulse profiles. The experimental photoelectron angular distributions were obtained
through subtracting the background single photon ionization harmonic peaks, however the
experimental pulse is much broader than the idealized theoretical pulse. The experimental
intensity profile of the 13$^{\textrm{th}}$ and 15$^{\textrm{th}}$ SPI harmonic peaks of
Argon in \cite{Haber09} show that there is a greater background signal present in the region of
the negative sidebands than the positive sidebands of each of these harmonics. This may affect
the extent to which theory and experiment can be compared. The narrow theoretical pulse ensures
that background subtraction is not necessary for the TDRM method.


For the MP calculations, we obtain a photoelectron angular distribution for each of the relevant sidebands
using the technique presented in \cite{Toma02}, before fitting Legendre polynomials in the form of
equation (\ref{eq:leg_fit}). As data for the positive sideband of the 19$^{\textrm{th}}$ harmonic
was not provided in \cite{Toma02}, these values are omitted in table \ref{tab:ArResults}.
The MP method is limited to considering a single active electron in the configuration space, thus
limiting its ability to describe multielectron effects such as correlation properly.
The MP method accounts for correlation and exchange effects through the use of a modified potential
chosen to reproduce the eigen energies and binding energies of the singly excited states of Argon.
It therefore excludes, for example, effects from the $3s3p^6nl/ \epsilon l$ channels. In order to
investigate how appropriate this potential is for quantitative studies, it is useful to investigate
how it compares to one that represents the Ar atom from first principles.

When compared with the anisotropy parameters obtained using the TDRM method, the MP approach provides
values for the anisotropy parameters that are in general agreement with those obtained in the TDRM
method. However, it can also be seen that the MP anisotropy parameters are consistently higher than
those obtained using TDRM: the \Btwom and \Bfourm parameters are approximately 0.2 higher than those
calculated using the TDRM approach, and similarly the \Btwop and \Bfourp are between 0.2-0.5 higher
than the TDRM equivalent. Since the TDRM asymmetry parameters are in all cases closer to experiment
than the MP parameters, it appears that the potential used in the TDRM calculations provides the
better approximation to the true Ar potential.


The second approach that we compare with is the soft-photon approximation.
In this approximation, the expression for the photoelectron angular distributions for two-color above
threshold ionization is given as \cite{Maquet07}
\begin{equation}
\label{eq:SPAfull}
\left(\frac{d\sigma^{(n)}}{d\theta}\right)_{E_k}=\frac{k}{k_0}J_n^2(\mathbf{\alpha_0\cdot K})
\left(\frac{d\sigma^{(0)}}{d\theta}\right),
\end{equation}
where $n$ is defined by the number of low-energy photons exchanged after absorbing a single high
energy photon, with the sign of $n$ determined by emission ($n>0$) or absorption ($n<0$) of the
low-energy IR photon. $n$ thus corresponds to the sideband in question (positive sidebands are
a result of absorption and negative sidebands are a result of stimulated emission).
$d\sigma^{(0)}/d\theta$ indicates the differential cross section for single photon
ionization given at the sideband energy $E_k$. The $\mathbf{\alpha}_0$ term represents
the classical excursion vector of a free electron in a laser field, while $\mathbf{K}$
is the momentum transfer between the electrons incoming wave vector and its final state wave vector.
$J_n$ represents a Bessel function. At low intensity, following \cite{Haber09}, the assumption is
made that $n=\pm1$ sidebands dominate and $|\alpha_0 \cdot K| \ll 1$. Consequently the Bessel
functions $J^2_{\pm1}(z)$ are proportional to $\cos^2\theta$, which results in the sideband
angular distributions given by equation (\ref{eq:SPAfull}) with $n=\pm1$ being determined
primarily by $\cos^2 \theta$ times the single photon differential cross sections. These
single photon differential cross sections behave as a function of theta as \cite{Maquet07}:
\begin{equation}
\label{eq:SinglePAD}
I(\theta)\propto 1+\beta_2 P_2(\cos \theta).
\end{equation}
We therefore obtain photoelectron angular distributions determined by the single
photon anisotropy parameter $\beta_2^{(0)}$. The value of $\beta_2^{(0)}$ is chosen
to correspond to the energy of each sideband. To enable a comparison between the
TDRM and SPA methods in two-color two-photon ATI, we choose to use $\beta_2^{(0)}$ parameters
calculated using the TDRM method as our input for the SPA method. This allows us to investigate
of the differences between the soft photon approximation and the TDRM approach with the least
influence from differences in the absorption of the harmonic photon. The values of $\beta_2^{(0)}$
are shown in table \ref{tab:TDRMbeta}. When considering the SPA results, we note that the values
we obtained differ from those used in \cite{Haber09} where experimental values
for $\beta_2$ were chosen.

\begin{table}
\caption{Single photon anisotropy parameters and photoionization cross sections
calculated using the TDRM method, and compared to the values used in \cite{Haber09}. The
latter data were originally obtained from \cite{Taylor77} and \cite{Henke93} respectively.}
\label{tab:TDRMbeta}
\begin{center}
\begin{tabular}{ccccc}
\hline
\hline
 Harmonic & TDRM & TDRM  & (\cite{Haber09}) & (\cite{Haber09}) \\
 & $\beta_2^{(0)}$ &  $\sigma$ (Mb) & $\beta_2^{(0)}$ &  $\sigma$ (Mb) \\
\hline
 12 & 0.4951 & 32.78 & 0.4 & 37.6\\
 14 & 0.9297 & 34.34 & 1.1 & 37.0\\
 16 & 1.2040 & 33.64 & 1.4 & 34.9\\
 18 & 1.3975 & 30.37 &     &     \\
 \hline
\hline
\end{tabular}
\end{center}
\end{table}

This protocol for the SPA method has several consequences for the angular distributions. First of all, the
 choice of the single-photon asymmetry parameter at the final-state energy means that the method predicts
no difference between the asymmetry parameters for the $N-1$ harmonic + IR absorption, and the $N+1$
harmonic + IR emission. This behaviour is seen in table \ref{tab:ArResults}. A second consequence is that
the SPA predicts a zero in the angular distribution perpendicular to the polarisation direction of the laser
fields, so that
\begin{equation}
\beta_4 = \frac{4}{3}\left(\beta_2 -2\right),
\end{equation}
as can also be seen from table \ref{tab:ArResults}.

Compared with the TDRM results, the anisotropy parameters for the SPA approach in table \ref{tab:ArResults}
are highly similar for positive sidebands with agreement between the two approaches improving with higher
energy harmonics. For negative sidebands however, the SPA approach provides anisotropy parameters that
are consistently higher than those obtained using TDRM ranging from 0.20 for the \Btwom parameter of the
19$^{\textrm{th}}$ harmonic to 0.45 for the \Btwom parameter of the 13$^{\textrm{th}}$ harmonic. The
effect of the atomic potential is therefore significantly more apparent for the negative sideband than for the
positive sideband. Compared with experimental results the SPA approach is similar to the TDRM approach for
the positive sideband of the 15$^{\textrm{th}}$ harmonic, and only marginally worse for the positive sideband
of the 13$^{\textrm{th}}$ harmonic. For negative sidebands the TDRM approach provided better approximations
to the experimental results than the SPA approach, although neither method lies entirely within the experimental
error bars. All of the values obtained using the SPA method lie outside the range of experimental error.

The overall comparison between the different approaches shows that the TDRM method provides anisotropy
parameters with a slightly better agreement with experiment than either the SPA or MP methods. The discrepancy
between the TDRM approach and the SPA and MP methods is demonstrated most notably for negative sidebands
of the 13$^{\textrm{th}}$ and 15$^{\textrm{th}}$ harmonics. It is not surprising that the main differences are
seen for these sidebands, since the atomic potential affects electron motion more for the lower harmonics than for
higher harmonics. Threshold effects also will be more influential for lower harmonics. As the order of the harmonic is
increased, the SPA and MP methods produce increasingly similar anisotropy parameters. However, they remain higher
than the TDRM parameters. The comparison of TDRM and SPA parameters is most sensitive to the description of the
continuum wavefunctions, since the ionization stage of the process is described by the TDRM method in both cases.
The comparison with the MP method, on the other hand, is significantly affected by the differences in the effective
potentials in the two methods during the ionization stage as well. This increases the potential for deviation between
the methods. This may account for the generally better agreement between the TDRM and the SPA approach than
between TDRM and MP.


The ratios between cross sections for positive and negative sidebands presented in table \ref{tab:ArResults}
indicate the relative strength of the absorption and emission of an IR photon (corresponding to a positive and
negative sideband respectively) for a given harmonic. The ratios obtained using the TDRM method demonstrate
that absorption process of an IR photon is the more likely process at all considered harmonics. However, the
sidebands tend towards equality at higher harmonics. These ratios obtained using the TDRM method are larger
than the experimentally obtained ratio by about 7-10\%, and lie outside the range of experimental error. As
indicated earlier, this may in part be due to differences between the theoretical model of the harmonic radiation
and the actual experimental frequency profile.

The cross section ratios predicted by the MP method for the 13\textsuperscript{th} and 15\textsuperscript{th} harmonics are in excellent agreement with experimental ratios, and slightly lower than those obtained using TDRM. The MP values display the same general downward trend with increasing energy as the TDRM values, however for the 17\textsuperscript{th} harmonic the ratio predicted by the MP method is higher than both the SPA and TDRM approaches. Despite
strong agreement with experimental values, there are noticeable differences in the asymmetry parameters, which suggests that the MP
approach does not describe the two-photon ionization process in full detail.


Within the SPA approach, the ratio between the cross sections for a positive and negative sideband within the
SPA is obtained by integrating equation (\ref{eq:SPAfull}) over all angles \cite{Haber09},
\begin{equation}
\label{eq:CSR_SPA}
\frac{\sigma^{(+)}}{\sigma^{(-)}} \propto \frac{k^{(+)}}{k^{(-)}}
\frac{\sigma^{(0,+)}}{\sigma^{(0,-)}}\frac{[5+2\beta_2^{(0)}]_{E_{k(+)}}}{[5+2\beta_2^{(0)}]_{E_{k(-)}}},
\end{equation}
where $\beta_2^{(0)}$ is the single photon anisotropy parameter from equation (\ref{eq:SinglePAD}) at
the energy $E_{k(\pm)}$ corresponding to the positive and negative sidebands for a selected harmonic,
$k^{(+)/(-)}$ indicates the momentum of the outgoing electron at the positive $(+)$ and
negative $(-)$ sideband, and $\sigma^{(0)}$ indicates the photoionization cross section at
the photon energy corresponding to the sideband energy. In order to determine this ratio, we have
obtained single-photon ionization yields at photon energies corresponding to the sideband energies, and
derived single-photon ionization cross sections from them. These cross sections are reported in table \ref{tab:TDRMbeta}.


The cross section ratio obtained using the SPA method for the 13\textsuperscript{th} harmonic is
significantly larger than both the TDRM and MP methods at this energy. For higher harmonics, the
SPA cross sections are found to be in very good agreement with the TDRM method. This suggests
that, as was the case with anisotropy parameters, at lower energies the atomic potential and
threshold effects, which are not included in the SPA method, have a significant effect on
sideband generation process. At higher energies these processes are less significant, thus the
SPA method predicts values that are largely similar to the TDRM method.
The ratio for the 15\textsuperscript{th} harmonic is in excellent agreement with
experiment and both the MP and TDRM approaches.

\section{Conclusions}

We have applied the TDRM method to calculate the anisotropy parameters for positive and negative sidebands of the
13$^{\textrm{th}}$ to 19$^{\textrm{th}}$ harmonics of an 800 nm pulse overlapped by an IR dressing field in two-color
two-photon above threshold ionization of argon and compared to those obtained experimentally, and by the MP and SPA
methods. Overall, the asymmetry parameters obtained by all three theoretical methods are in general agreement with
each other. At a detailed level, the anisotropy parameters calculated using the TDRM method are found to be generally
smaller than those obtained using the SPA and MP approaches. Of the SPA and MP methods, the SPA approach provided
anisotropy parameters closer to those predicted by TDRM method and measured experimentally for positive sidebands,
but also produced values with the lowest degree of agreement with experiment for negative sidebands. The MP method
provides values closer to TDRM and experiment than SPA for negative sidebands. For higher photon energies the differences
between the TDRM method and the SPA and MP methods decrease.


The anisotropy parameters calculated using the TDRM method are in good agreement with those predicted experimentally
and show an improvement in agreement with experimental results when compared to anisotropy parameters obtained using
the MP and SPA methods. This demonstrates that the TDRM method gives some improvement in theoretical modeling of
two-color two-photon above threshold ionization of Argon relative to the MP and SPA methods. This improvement probably
originates from a more accurate description of the potential seen by the outer electron. While the TDRM method shows
better agreement with experiment than either the SPA or MP methods, the anisotropy parameters obtained still remain
mostly outside of the range of experimental values. Some of these differences may be due to differences between the
frequency profile used in the theoretical calculations for the harmonic laser pulse and the experimental frequency profile.

\section*{Acknowledgments}
SH was supported by the Department of Employment and
Learning NI under the Programme for Government.
HWvdH is supported by the UK Engineering and Physical Sciences Research Council under
grant number G/055416/1.

\bibliography{arpaper}

\end{document}